\documentclass[sigplan,10pt]{acmart}

\copyrightyear{2026}
\acmYear{2026}
\setcopyright{cc}
\setcctype{by}
\acmConference[EUROSYS '26]{21st European Conference on Computer Systems}{April 27--30, 2026}{Edinburgh, Scotland Uk}
\acmBooktitle{21st European Conference on Computer Systems (EUROSYS '26), April 27--30, 2026, Edinburgh, Scotland Uk}
\acmDOI{10.1145/3767295.3803576}
\acmISBN{979-8-4007-2212-7/2026/04}

\usepackage{pifont}
\usepackage[T1]{fontenc}
\usepackage{zi4} 
\usepackage{multirow} 
\usepackage{enumitem}
\newlist{tightitemize}{itemize}{1}
\setlist[tightitemize]{leftmargin=*,label=\textbullet,nosep,topsep=0pt}
\newlist{tightdesc}{description}{1}
\setlist[tightdesc]{style=unboxed,format=\bfseries, leftmargin=1.6em,labelsep=0.35em,nosep,topsep=0pt}
\usepackage{tabularx,makecell,booktabs}
\usepackage{marvosym}
\newcommand{\heading}[1]{
\vspace{1ex}
\noindent
\textbf{#1}
}

\def\compactify{\itemsep=0in \topsep=2pt \parsep=0.00in \partopsep=0pt
\leftmargin=1.4em}
\let\latexusecounter=\usecounter

\newenvironment{myitemize}%
  {\begin{list}{\labelitemi}{\itemsep1pt \topsep2pt \parsep0.00in
  \partopsep=0pt \leftmargin1.4em}}%
  {\end{list}}

\usepackage{amsmath}

\usepackage{svg}

\usepackage{xspace}
\usepackage{listings}
\lstdefinestyle{tightjson}{
  basicstyle=\ttfamily,    
  columns=fullflexible,
  showstringspaces=false,
  xleftmargin=1em,          
  aboveskip=2pt, belowskip=2pt
}

\usepackage{graphicx}
\usepackage{subcaption} 
\usepackage{balance}



\newcommand{\sys}{\textsc{DMI}\xspace}
\begin{document}

\title[]{From Imperative to Declarative: Towards LLM-friendly OS Interfaces for Boosted Computer-Use Agents}

\author{
{Yuan Wang$^{\star\diamondsuit\dagger}$ \enspace Mingyu Li$^{\star\diamondsuit}$ \enspace Haibo Chen$^{\star\diamondsuit\ddagger}$}
}
\affiliation{%
  \institution{$^\star$Key Laboratory of System Software (Chinese Academy of Sciences)}
  \city{}
  \country{}
}
\affiliation{%
  \institution{$^\diamondsuit$Institute of Software Chinese Academy of Sciences}
  \city{}
  \country{}
}

\affiliation{%
  \institution{$^\ddagger$Shanghai Jiao Tong University}
  \city{}
  \country{}
}

\affiliation{%
  \institution{$^\dagger$University of Chinese Academy of Sciences}
  \city{}
  \country{}
}

\begin{abstract}

Computer-use agents (CUAs) powered by large language
models (LLMs) have emerged as a promising approach to automating computer tasks, yet they struggle with the existing human-oriented OS interfaces---graphical user interfaces (GUIs). GUIs force LLMs to decompose high-level goals into lengthy, error-prone sequences of fine-grained actions, resulting in low success rates and an excessive number of LLM calls.

We propose Declarative Model Interface (\sys), an abstraction that transforms existing GUIs into three declarative primitives: access, state, and observation, thereby providing novel OS interfaces tailored for LLM agents. Our key idea is policy-mechanism separation: LLMs focus on high-level semantic planning (policy) while \sys handles low-level navigation and interaction (mechanism). \sys does not require modifying the application source code or relying on application programming interfaces (APIs).

We evaluate \sys with Microsoft Office Suite (Word, PowerPoint, Excel) on Windows. Integrating \sys into a leading GUI-based agent baseline improves task success rates by 67\% and reduces interaction steps by 43.5\%. Notably, \sys completes over 61\% of successful tasks with a single LLM call.

\end{abstract}



\maketitle

\section{Introduction}
Computer-use agents (CUAs) powered by large language models (LLMs) demonstrate remarkable potential in automating complex workflows, creating unprecedented opportunities for productivity enhancement. Recent breakthroughs have catalyzed intense interest across both industry and academia \cite{ClaudeComputerUse,OpenAIOperator,osworld,ufo2,uitars,uivground,screenagent, mobileagentv3, OSWorld-Human, WebVoyager, AutoWebGLM}. To interact with applications running on computer systems, state-of-the-art CUAs primarily rely on two interfaces: application programming interfaces (APIs) and graphical user interfaces (GUIs). 

Under the first paradigm, API-based approaches capitalize on the code-generation capabilities of LLMs to directly invoke application-specific APIs~\cite{apifirst,ufo2,ComputerRL, apillmssurvey}. While this approach typically yields higher success rates and requires fewer execution steps, it faces a critical limitation: many applications lack publicly available APIs, thereby restricting the approach's overall generality. On the other hand, GUI-based approaches~\cite{OpenAIOperator, ClaudeComputerUse, uitars, ufo,ufo2} leverage LLMs (including multimodal LLMs) to perceive screen content, locate interface elements, and emulate human actions such as clicks and scrolls. This approach offers unparalleled generality since GUIs represent the dominant human-computer interface paradigm on modern desktop and mobile operating systems. However, it demands that LLMs generate lengthy, fine-grained action sequences to manipulate UI control elements (hereafter, controls), increasing LLM inference costs while compounding the risk of cascading failures~\cite{apifirst,apivsgui,gta1,selfcorrect,LLM-GUI-Agent-Survey}.

Our investigation suggests that the need for complex action sequences in GUI use stems from the \textbf{imperative} nature of GUI design. First, unlike APIs with direct function calls, GUIs require navigating through spatial layouts to reveal hidden or off-screen controls. This forces the LLM to execute chains of \textbf{navigation} actions (e.g., navigating menus, tabs, and dropdowns) to make the control visible.
Second, GUI controls require explicit \textbf{interaction} (e.g., a click) to trigger a function. Beyond simple clicks, many controls demand composite interactions. For instance, manipulating a scrollbar requires a coordinated sequence of actions: moving the cursor to the scrollbar, pressing the
mouse button, dragging to a target position while holding
the button, and finally releasing it.

This imperative paradigm creates an entanglement of two distinct aspects in GUI application use.
The first is the orchestration of application functionality according to task semantics, which we term \textbf{policy}. The second is the process of using functional controls, which involves navigation and interaction, an aspect we refer to as \textbf{mechanism}. 

In current GUIs, policy is tightly coupled with mechanism:
users cannot directly invoke the function; they must perform the requisite navigation and interaction. This coupling poses a significant bottleneck for LLMs; it substantially increases task complexity, strains the LLM's spatial reasoning and visual processing, and necessitates a higher number of LLM invocations, inflating both token costs and latency.

This raises an essential research question: \emph{How should OSes evolve interfaces to serve a fundamentally new class of OS users (LLMs), instead of forcing them to mimic humans and use existing human-oriented interfaces (namely, GUIs) for which they are ill-suited?}

Our key insight is that \emph{while LLMs struggle with fine-grained mechanisms, a substantial fraction of the interaction logic with GUI is deterministic and can be executed independently of the LLM.} Based on this observation, we introduce Declarative Model Interface (\sys), an abstraction layer that transforms complex GUI navigation and interaction into three declarative primitives: \textit{\textbf{access}}, \textit{\textbf{state}}, and \textit{\textbf{observation}}. By handling navigation and interaction deterministically, \sys decouples policy from mechanism---a separation of concerns that shifts the LLM's role from ``orchestrate both high-level functions and low-level UI actions'' to ``orchestrate only the semantic, non-deterministic aspects of the task.''

\begin{figure}[t]
\centering 
\includegraphics[width=0.85\columnwidth]{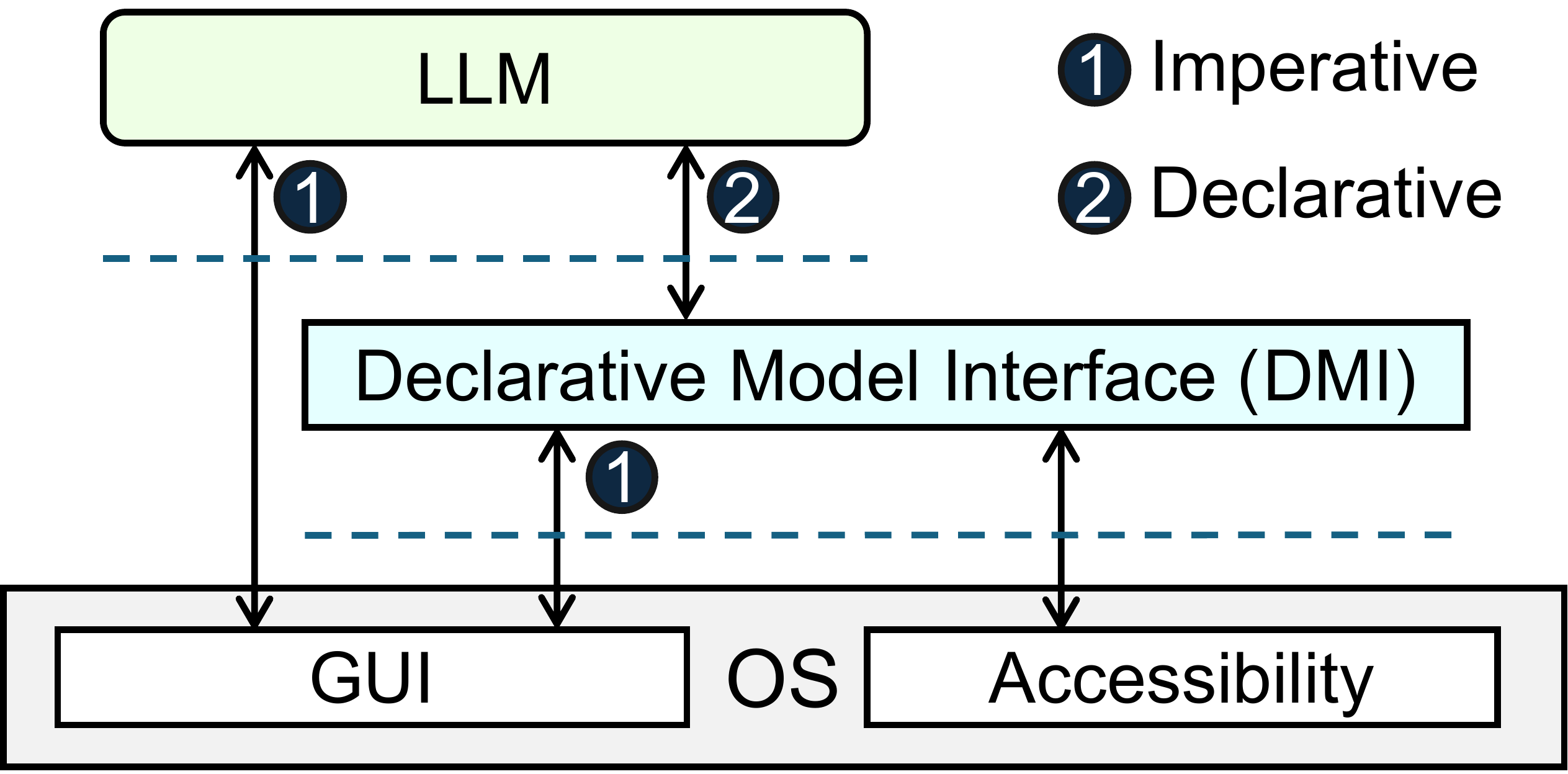}
\caption
{\small Overview of the \sys abstraction layer.
 \sys is based on ubiquitous GUI and OS accessibility features~\cite{OSAccessibility, WindowsUIA, Linux-AT-SPI2}. 
\textit{Declarative} specifies the intended state or outcome;
\textit{imperative} enumerates the actions that realize it. \sys abstraction provides novel \textbf{OS interfaces} tailored for LLM agents.}

\label{interface}
\end{figure}

Unlike imperative GUI interaction, declarative \sys allows and requires the LLM to specify the desired outcome directly, rather than emitting concrete actions to realize the outcome. Using {\sys}'s declarative primitives, the LLM can express the intended result for a control: 
the control is to be ``accessed'' or ``set to a target state'', or to have its information ``observed and retrieved''. \sys then executes the necessary steps to achieve that result. This procedure is fully transparent to the LLM. Such a design meets our goal: to let the LLM abstract away from concrete GUI operations, and eventually focus on semantic planning for the user's task.

There are several challenges to be addressed to realize \sys.
First, the controls' navigation relationships are typically implicit within the applications, which obstructs navigation abstraction. 
Second, a control's functionality is path-dependent. Take Microsoft Word as an example:
accessing a standard color control via the \texttt{Font}, \texttt{Outline}, or \texttt{Underline} path can yield different semantics. 
Enforcing path uniqueness through duplication (i.e., cloning the merge node and its substructure)  can cause an exponential blow-up in an already large application node set, which in turn highlights another major issue: LLM context is limited and costly. 
\sys design must account for invocation cost. 
Additionally, the ideal interface should provide a stable abstraction: this requires \sys to handle the non-determinism of GUI execution robustly. 
Finally, an LLM-friendly \sys must also consider the caller's (the LLM's) imperfect instruction-following. 
For example, while we explicitly instruct the LLM to output target function controls only, the LLM may still output navigation paths.

We present an end-to-end solution to these challenges. Specifically, we model controls' navigation relationships and transform the resulting  graph into a path-unambiguous topology (\autoref{sec:Path-Unambiguous}). 
We introduce an LLM-context-friendly representation of controls and navigation, and design an on-demand querying mechanism to balance context cost, function coverage, and efficiency (\autoref{sec:Context-Efficient}). \sys is designed with an ``LLM-as-user'' assumption; to mitigate imperfect instruction-following, we apply filtering to ensure \sys entirely takes over the navigation process. To provide a stable declarative abstraction, \sys incorporates robustness mechanisms to handle instability in executing UI action sequences (\autoref{sec:visit}). We enforce a strict separation between control access and complex interactions to relieve the tension between accuracy and efficiency (\autoref{sec:visit}, \autoref{sec:InteractOps}).

We validate \sys through a set of case studies against Microsoft Word, Excel, and PowerPoint. 
These applications cover a diverse range of real-world scenarios, including text editing, spreadsheet manipulation, and graphics processing. 
Our evaluation with the OSWorld-W \cite{osworld} dataset shows that \sys outperforms UFO-2 \cite{ufo2}, the leading GUI-based baseline, by increasing the absolute task success rate by 29.6\% (67\% relative improvement), reducing interaction steps by 43.5\%, and decreasing completion time by 39\%. 
Regarding failure cases, the UFO-2 baseline exhibits numerous failures related to the mechanism, such as errors in visual recognition, fine-grained interaction, and navigation.
In contrast, with \sys, over 80.9\% of failures are policy-related (e.g., LLM semantic misunderstandings), rather than mechanism-related. These results demonstrate the effectiveness of declarative interfaces as LLM-friendly interaction paradigms.

In this paper, we make the following contributions:
\begin{myitemize}
\item We identify that human-oriented OS interfaces (i.e., GUIs) pose challenges for LLM-driven CUAs, a fundamentally new class of OS users with distinct characteristics and constraints.
\item We introduce \sys, an abstraction that decouples policy from mechanism by reducing complex GUI operations to declarative primitives, enabling LLMs to focus on semantic planning. \sys provides LLM-friendly OS interfaces.    
\item We conduct an extensive evaluation showing that \sys significantly improves LLM task success rates and efficiency compared to state-of-the-art GUI-based approaches.
\end{myitemize}

\section{From Imperative to Declarative}
\subsection{Human-centric Design: A Hurdle for LLMs}

We first examine the paradox of GUIs: why the human-centric OS interface design, which is intuitive for humans, creates significant obstacles for LLMs.

\heading{Mismatch \#1: Procedural \emph{navigation}.}
GUI applications expose functionality in a step-by-step manner: controls are hierarchically organized behind menus, tabs, and dialog boxes. This design deliberately narrows the set of available choices at each step, decomposing control localization into low-cognitive-load decisions \cite{MicrosoftInterfaceDesign}, while favoring visual recognition over memory-based recall \cite{recognition_over_recall}.
The design assumes users (i) struggle with large decision spaces due to limited working-memory capacity, (ii) have poor exact syntax recall (e.g., command names, argument order), but (iii) excel at visual recognition. 
These assumptions do not hold for LLMs. LLMs can leverage a much larger context window (e.g., GPT-5, 400K tokens) and reliably produce grammar-constrained, structured outputs \cite{OpenAIStructuredOutputs, ClaudeToolUse}, but they have comparatively weaker visual perception \cite{SpiritSight,GUIagent,LoRALMM}. Given a global view, whether from trained knowledge or task-specific documentation in the prompt, an LLM can directly identify the appropriate control and generate structured invocations \cite{Gorilla}. However, an imperative GUI requires LLMs to first produce navigation sequences that make the target controls visible. This increases the length of action chains and introduces systemic fragility, where any planning or execution error can cascade into complete task failure.

\heading{Mismatch \#2: Iterative \emph{interaction}.}
Many GUI controls rely on iterative interaction. For example, text selection involves repeated cursor movements to establish start and end positions. This design enforces high-frequency ``observe–act'' loops that transform challenging, high-precision outputs (exact coordinates) into manageable visual judgments (``is the current state acceptable?''). This assumes (i) frequent ``observe–act'' loops incur minimal cost, and (ii) perception is effortless and accurate.
Neither assumption holds for LLMs. LLM inference latency makes frequent closed-loop control prohibitively expensive: while humans perform 3–5 mouse adjustments per second with real-time feedback, LLM inference requires 10–120+ seconds per round-trip \cite{llminferencetime,serverlessllm}. In addition, LLMs exhibit limited visual acuity, making precise UI understanding unreliable \cite{SpiritSight,GUIagent,LoRALMM}.

\begin{table}[t]
\captionsetup{skip=3pt}  
\caption{\small Task examples of imperative GUI vs. declarative \sys.}
\label{tab:GUIvsDMI}
\centering
\scalebox{0.65}{
\begin{tabular}{cll}
\toprule
\textbf{Task} & \makecell[c]{\textbf{GUI}} & \makecell[c]{\textbf{\sys}} \\
\midrule
1 & \makecell[l]{\texttt{click}("{Design}") $\to$ \texttt{click}("{Format Background}") \\ $\to$ \texttt{click}("{Solid fill}") $\to$ \texttt{click}("{Fill Color}") \\ $\to$ \texttt{click}("{Blue}") $\to$ \texttt{click}("{Apply to All}")} & \texttt{visit}(["{Blue}", "{Apply to All}"])
\\
\midrule
2 & iterative interaction (\texttt{drag} and \texttt{drop}) & \texttt{set\_scrollbar\_pos({80\%})}
\\
\bottomrule
\end{tabular}
}
\end{table}

\begin{figure}[t]
  \centering
  \begin{subfigure}{0.9\linewidth}
    \centering
    \includegraphics[width=\linewidth,trim=2mm 3.5mm 0mm 0mm,clip]{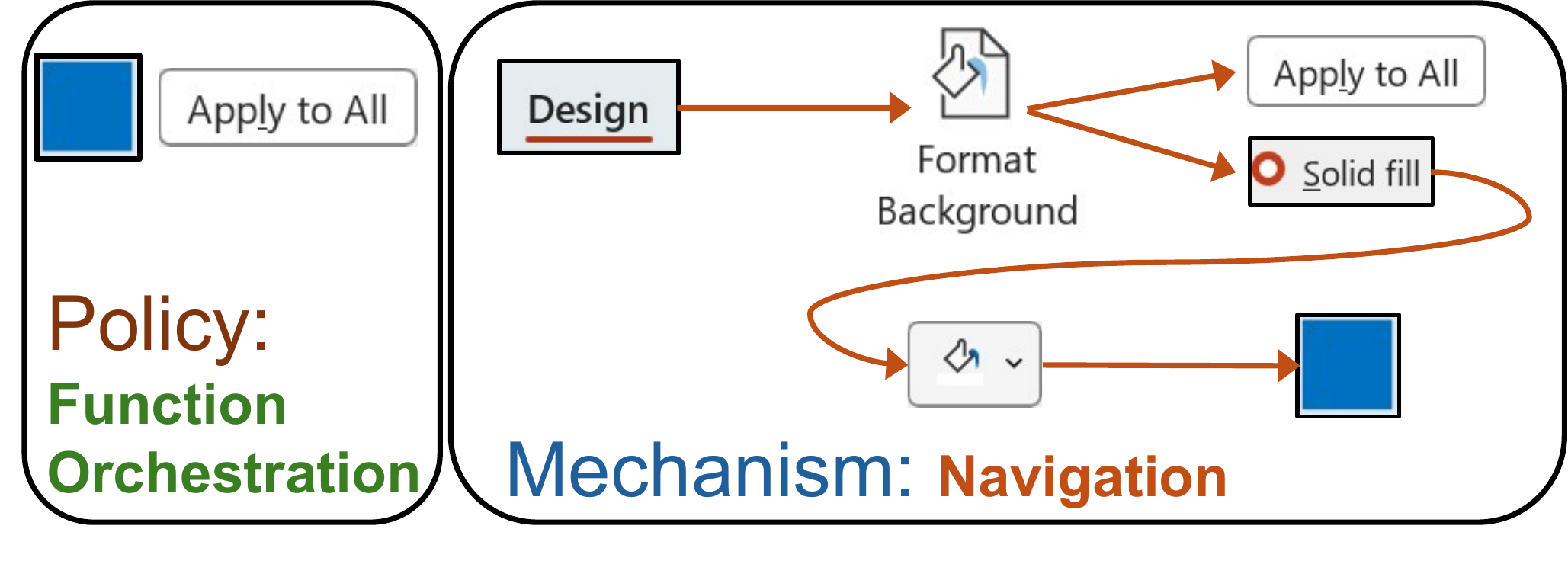}
    \caption{Task 1: make the background blue on all slides.}
    \label{fig:mechanism_policysub1}
  \end{subfigure}
  \begin{subfigure}{0.9\linewidth}
    \centering
    \includegraphics[width=\linewidth,trim=2mm 4.5mm 0mm 0mm,clip]{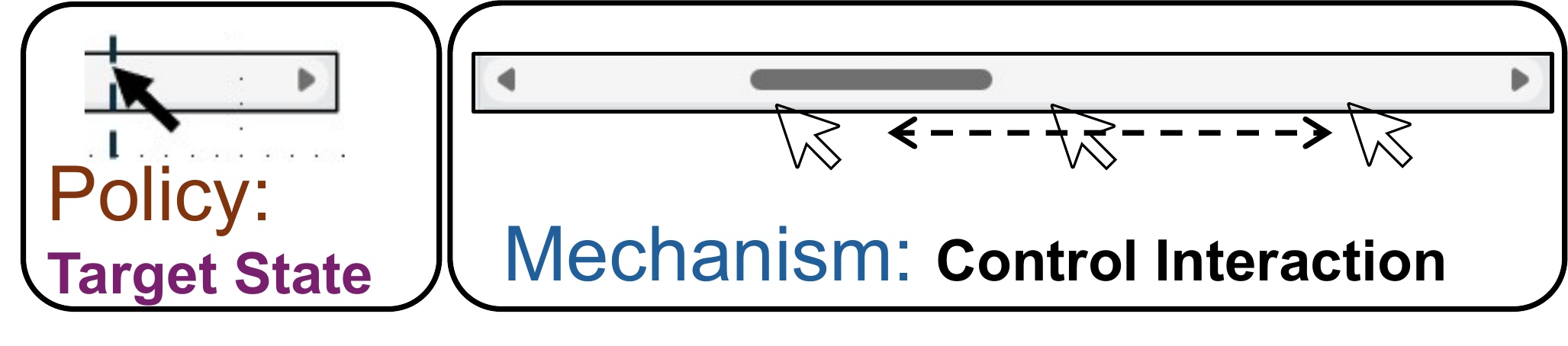}
    \caption{Task 2: show the area close to the end.}
    \label{fig:mechanism_policysub2}
  \end{subfigure}

\caption{\small Policy-Mechanism coupling in GUI use.}
\label{fig:mechanism_policy}

\end{figure}

\subsection{Insights}

\heading{Insight \#1: Policy–mechanism decoupling.}
When using GUI applications, \textbf{policy} (function orchestration) becomes tightly coupled with the mechanism (control navigation and interaction). \autoref{fig:mechanism_policy} and \autoref{tab:GUIvsDMI} illustrate this coupling through two real-world examples. Since human-centered GUIs misalign with LLM capabilities, LLMs perform poorly on mechanism-level tasks.
In Task 1, invoking ``Blue'' requires emitting a precise five-step navigation sequence; any single misstep could invalidate the plan. In Task 2, the LLM must perform the iterative drag–observe cycle on the scrollbar and rely on visual feedback, while the LLM is prone to misperceiving on-screen positions. Additionally, many LLM calls increase latency. However, much of this mechanism is \textbf{deterministic} and can be resolved algorithmically without LLM involvement. Rather than forcing LLMs to plan fragile navigation sequences and complex interactions, we propose offloading deterministically solvable mechanisms to an abstraction layer, thus enabling LLMs to focus on nondeterministic, policy-level decisions that require semantic reasoning. 
 
\heading{Insight \#2: Deterministic navigation.}
Given a target control, computing an access path and navigating to it becomes a deterministic problem. Once an application is released, its control transitions form a finite state machine \cite{autodroidv2}, with control reachability and dependencies modeled as a directed graph where nodes represent controls. While controls may be accessible through multiple paths, removing back-edges and duplicate nodes with their successors yields a tree structure, producing a \textbf{unique} path identifiable from the control's identifier alone. This approach eliminates navigation dependencies from GUI use and compresses lengthy action chains. The LLM needs only to declare the target control rather than specify the concrete navigation sequence, removing responsibility for path correctness from the model.

\heading{Insight \#3: Finite interaction operations.}
While controls exhibit diverse behaviors, Windows UI Automation (UIA) categorizes them into finite sets: 41 control types (e.g., \texttt{Button}, \texttt{ListItem}, \texttt{Edit}) and 34 control patterns (e.g., \texttt{TextPattern}, \texttt{ScrollPattern}, \texttt{SelectionPattern}) \cite{MicrosoftControlPattern}.
These universals enable us to abstract low-level, fine-grained interactions into state and result declarations such as \texttt{set\_scrollbar\_pos}\allowbreak(x\_percent, \allowbreak y\_percent) or \texttt{select\_lines}(start\_index, \allowbreak end\_index). With these wrapped primitives, LLMs only need to specify the desired control state, eliminating requirements for precise visual-coordinate reasoning or high-frequency interaction.

\subsection{LLM-friendly Paradigm: Declarative Interfaces}

Decoupling \textbf{policy} (what to do) from \textbf{mechanism} (how to do it) allows for a shift from an \textbf{imperative} to a \textbf{declarative} interface paradigm. In this paradigm, an interface user (in this case, an LLM) can specify a desired state rather than planning and executing a long sequence of imperative actions. For example, the LLM could simply declare its desired outcome, such as ``set the \textit{scrollbar} position to 50\%'', or ``access the \textit{Apply to All} control''. The interface then handles the complex, low-level execution, regardless of the control's current state or visibility on the screen. \autoref{fig:mechanism_policy} illustrates this paradigm shift.

Declarative interfaces are well-suited for LLMs for two main reasons. 
First, this approach shifts interaction from state-based ``observe-act'' loops to target state setting.
This is crucial because it eliminates the dependency on the LLM's relatively weaker
precise visual perception and minimizes the need for high-frequency, real-time operations. Instead of constantly analyzing the screen to decide the next step, the LLM simply states the end goal.
Second, by hiding the low-level, fine-grained details of interaction, this paradigm distills tasks down to their semantic reasoning, which enables LLMs to focus on their strengths: understanding high-level intent and emitting well-formed, grammar-constrained outputs.

\begin{figure*}[t]
\centering 
\scalebox{1}[1]{%
\includegraphics[width=\textwidth]
{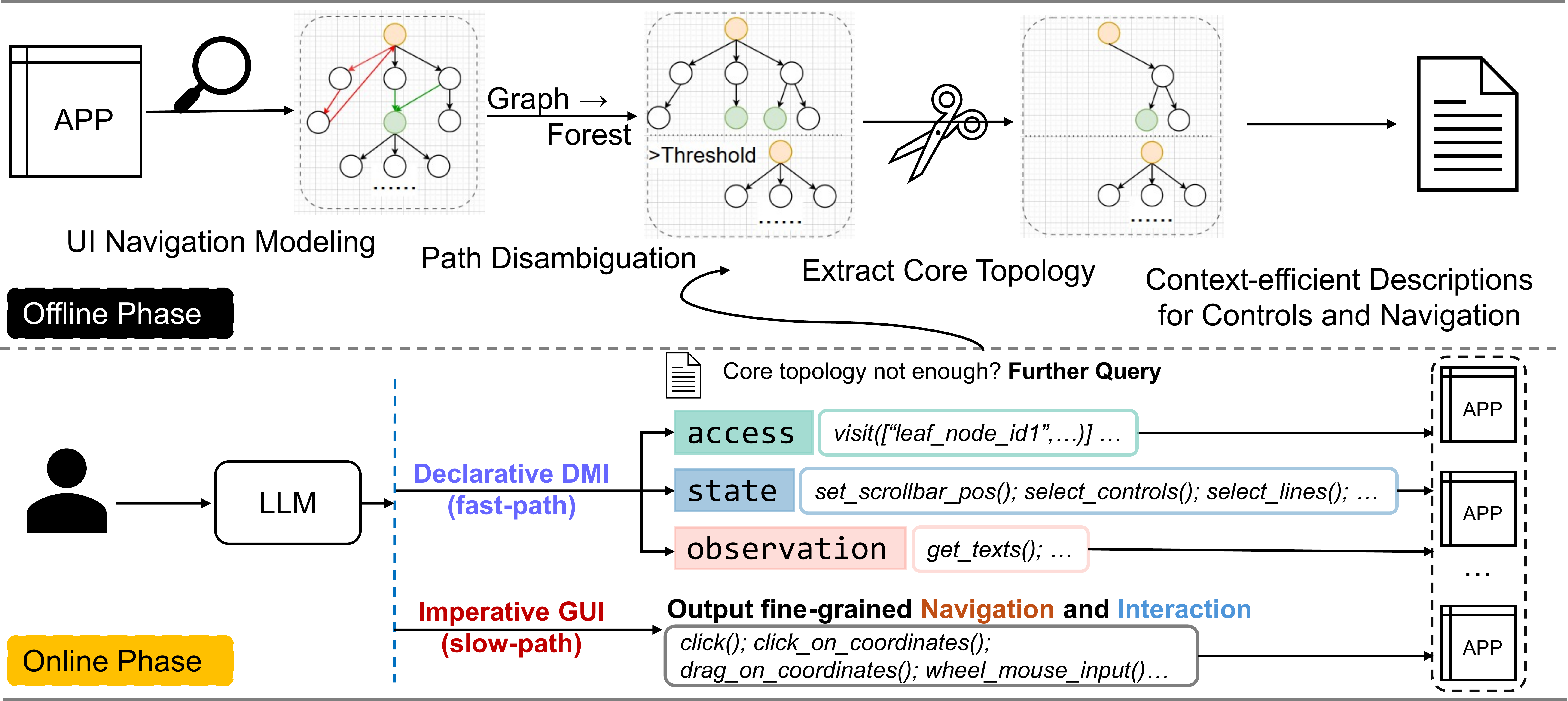} 
}
\caption{\small \sys Workflow: Offline Modeling and Online Execution. \sys has three declarative primitives: Access, State, and Observation.}
\label{fig:architecture}
\end{figure*}

\subsection{Challenges}
\label{sec:challenges}
To realize the shift from imperative to declarative interaction, we need to address the following challenges:

\heading{Challenge \#1: Navigation path ambiguity.}
Declarative control access requires explicit application navigation pathways, but these relationships exist implicitly and are not exposed as explicit data structures.
Even with deterministic UI topology, \emph{path ambiguity} exists: cycles and merge nodes in the navigation graph prevent mapping controls to unique access paths.

\emph{Cycles:} Loops (e.g., A → B → A) may yield infinite traversal sequences.

\emph{Merge nodes:} Controls may be reachable via multiple paths (e.g., A → C and B → C).

Identical controls can enact different functions depending on path-related context. For instance, in Word's color picker, a color cell is reachable via ``Font Color,'' ``Outline Color,'' or ``Underline Color''; the chosen path determines which property is modified.

\heading{Challenge \#2: Limited LLM context windows.}
Navigation topology should be converted to textual prompts for LLM comprehension. However,
modern applications expose numerous controls (>5,000 in Microsoft Office), making full topology prohibitively expensive for LLMs.
An ideal interface should enable LLMs to accurately identify required controls without overwhelming their context windows.

\heading{Challenge \#3: Inaccurate long-horizon planning.}
Since the navigation topology is deterministic, the LLM can plan multiple commands in a single call, even when target controls are not currently visible. However, this makes subsequent commands contingent on earlier correctness: unexpected intermediate outcomes can invalidate subsequent steps and cascade into task errors.

Additionally, real-world UI interaction is inherently unstable; for example, control name variations can break element identification.

\section{The \sys Design}
\label{sec:designsec}

\subsection{Overview}

The GUI mechanism includes control navigation and interaction. \sys abstracts navigation as \textit{access} declaration, and abstracts interaction as \textit{state} and \textit{observation} declarations.

\label{sec:designoverview}

\begin{myitemize}
\item \textit{Access declaration:}
Given a control identifier, \sys deterministically navigates from any current state to that control and performs a primitive interaction (e.g., click). 

\item \textit{State declaration:}
Given a desired control end state (e.g., scrollbar position; selection state for a control or for text), \sys transitions the control from any current state to the target state, abstracting away compound interactions such as drag and keyboard–mouse coordination.
\item \textit{Observation declaration:}
Given an information request (e.g., a control's text content), \sys returns structured data rather than relying on pixel-level recognition, without requiring compound interactions to reveal hidden content (e.g., expanding a table item).
\end{myitemize}

\heading{Interfaces:}
\sys provides two categories of interfaces. The \texttt{visit} interface is designed to take over navigation and realize \textit{access} declaration.
Additionally, \sys presents a set of interaction-related interfaces to support \textit{state} and \textit{observation} declaration. These interfaces simplify complex interactions such as scrolling, selection, and text manipulation into direct state setting and structured information retrieval.

\heading{Workflow:}
The workflow of \sys is shown in~\autoref{fig:architecture}. First, navigation relationships among controls are modeled as a graph, which is further translated into a path-unambiguous forest containing a main tree and shared subtrees. Next, this forest topology is converted into context-efficient textual representations with query-on-demand mechanisms to reduce token overhead. Finally, LLM is provided with the declarative interface to use intended controls.

\heading{Outline:}
\autoref{sec:Path-Unambiguous} introduces the navigation topology addressing path ambiguity (Challenge \#1). \autoref{sec:Context-Efficient} presents context-efficient descriptions for managing the LLM context window (Challenge \#2). \autoref{sec:visit} and  \autoref{sec:InteractOps} detail interface design for robust and efficient execution (Challenges \#3).

\subsection{Path-Unambiguous Navigation Topology}
\label{sec:Path-Unambiguous}
This section presents the application navigation modeling and how to construct the path-unambiguous topology.

\heading{Navigation modeling:}
Inspired by prior work \cite{guiripping,regui}, we build an automated prototype to model UI navigation relationships, with modest human intervention (detailed in \autoref{sec:navmodel}). The result is a UI Navigation Graph (UNG): nodes are UI controls and directed edges denote ``click'' interaction. 

UNG is a directed graph
$\mathcal{G}=(\mathcal{V},\mathcal{E})$ where each node $v\in\mathcal{V}$ corresponds to a UI control exposed by the accessibility API, and each edge captures \textbf{click}-induced reachability between controls.
In UNG, we only model control-to-control transition relationships; we do not include keyboard-shortcut actions as transition edges, since their effects are typically achievable via equivalent click operations.

\begin{figure}[t]
  \centering 
  
  \includegraphics[width=\columnwidth]{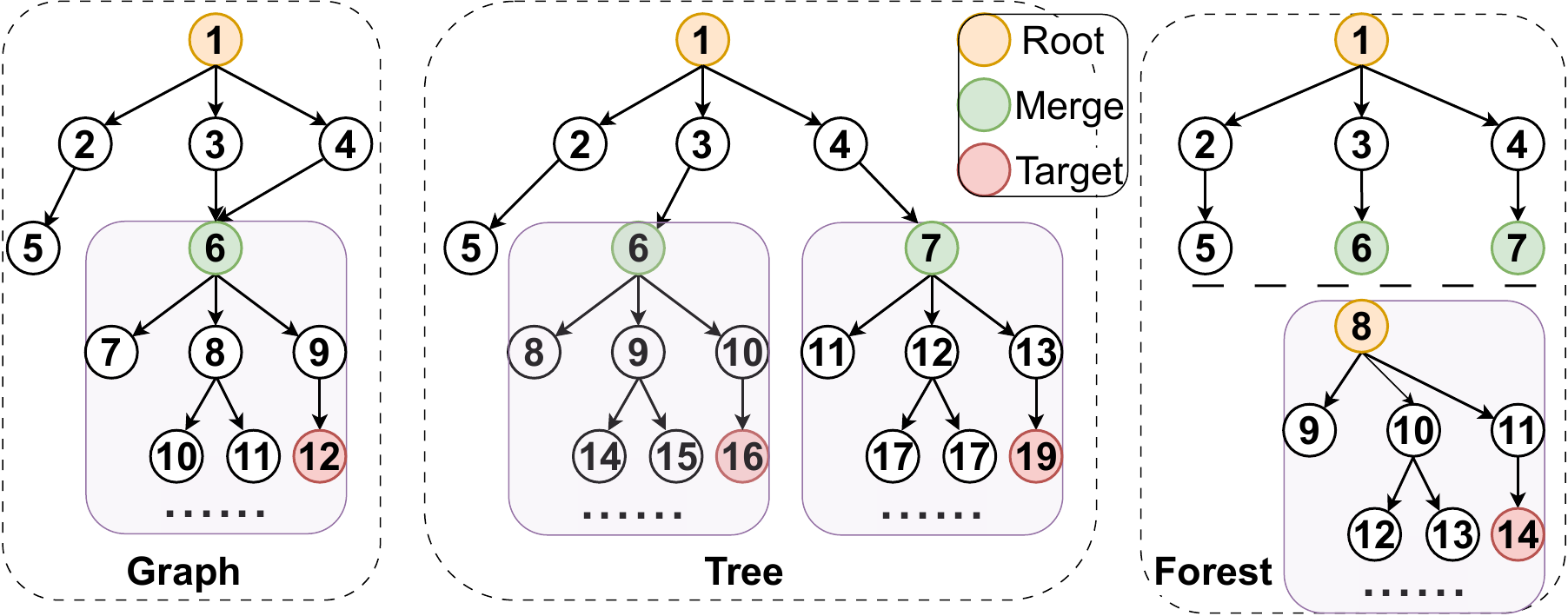} 
  \caption{\small Navigation Topology. To access the bottom-right node along node 4, imperative GUI navigation relies on graph and requires the explicit path $\langle1, 4, 6, 9, 12\rangle$. With declaration, only $\langle19\rangle$ is required by tree, while the number of nodes explodes. For forest, the path required is $\langle7, 14\rangle$.}
  \label{fig:forest}
\end{figure}

\heading{From directed graph to forest:}
To ensure unambiguous control access by declaring the control ID, UNG must be transformed into an \emph{ambiguity-free} topology. We define path ambiguity-free as follows: for any given control, a unique access path can be determined within its connected topology. The resulting structure is a forest, comprising a main tree and a set of shared subtrees. This transformation involves two critical steps.
First, we decycle the graph to a DAG. This process begins by removing cycles from the single-source UNG to produce a single-source DAG. This is achieved by identifying and removing back-edges that form the cycles.

The next step is to solve path disambiguation by turning the DAG into a forest, which focuses on handling merge nodes in the DAG (nodes with multiple incoming edges).
A naive approach would be to convert a merge node into a single tree by cloning the node and its entire descendant substructure for every incoming edge. While this guarantees unique paths, it causes an exponential node blow-up.
For complex applications (e.g., Microsoft Office), the resulting description significantly increases the context size, which exceeds the available context window of the LLM (400K tokens, GPT-5) in our experiment.
As discussed, simply deleting in-edges to enforce uniqueness is not an option, because different paths to the same GUI control can carry distinct semantics.

We designed a cost-based selective externalization algorithm to balance output-path length and total node count by transforming the single-source DAG into a forest. This bottom-up algorithm processes nodes in reverse topological order. For each merge node, it estimates the rooted substructure size and cloning cost—additional nodes from duplicating the substructure along all incoming edges. When this cost exceeds a configurable threshold, the node and descendants are externalized as a shared subtree, with incoming edges redirected to new reference nodes for indirect access. Otherwise, the substructure is cloned along each edge.
This approach ensures linear node growth.
In practice, the LLM specifies only a target control ID (see \autoref{fig:forest}) and reference IDs (typically one) for controls in shared subtrees. The executor then deterministically resolves navigation with reduced context overhead and bounded topology size while preserving semantic correctness and avoiding path ambiguity.

\subsection{Context-Efficient Descriptions for Controls and Navigation}
\label{sec:Context-Efficient}

This section explains how \sys converts the navigation topology into a textual format optimized for LLMs.

\heading{Strawman:}
A straightforward approach would discard navigation controls and abstract functional controls (the leaf nodes of the topology) into a flattened, API-like list of endpoints. However, this approach sacrifices critical semantic information. Many UI controls share generic names (e.g., ``Color,'' ``Settings,'' or ``OK'' buttons), and a flattened list introduces ambiguity that increases the risk of incorrect LLM selections. To resolve this ambiguity within a flat structure, one would need to encode the navigation path or its semantic meaning into each leaf node's description. This leads to massive data redundancy, as sibling nodes would share identical ancestor path information.

\heading{Compressed, hierarchical description:}
We encode the navigation hierarchy as compact, structured text. We retain the non-leaf nodes (navigation controls) as an integral part of each control's description. A control is thus described by a combination of its own properties and its hierarchical path. This keeps the organization of functionality explicit, allowing the LLM to disambiguate by full path, perform precise semantic reasoning, and avoid the duplication that a flattened representation would impose.
\sys relies on a shared subtree entry map to connect the main tree and shared subtrees.
This map records each reference node and the root of the subtree it targets. 
With this map and topology, the LLM can use \texttt{ref\_id} when locating controls within shared subtrees, ensuring a unique and correct navigation.

\heading{Query on demand:}
To conserve costly LLM context windows, we adopt a layered retrieval strategy based on the observation that most tasks require only partial navigation topology.
First, \sys by default provides a limited-depth core (e.g., six levels) instead of the complete forest, excluding large enumerations (font lists) and manually identified nodes. While pruning rules are currently manual, future versions could leverage LLM assistance for automation.
Second, when the pruned core lacks required structure, the LLM requests additional content via \texttt{further\_query} commands supporting two modes: (a) targeted branch queries expanding sub-structures beneath specified nodes, and (b) global queries retrieving the complete forest.
This design minimizes initial context load for routine tasks while enabling access to additional structures through targeted expansion or complete forest retrieval when necessary.

\subsection{Access Declaration: The \texttt{visit} Interface}
\label{sec:visit}

For \textit{access} declaration, the \texttt{visit} interface enables callers to directly access target controls via identifiers rather than navigation sequences. The \texttt{visit} interface also encompasses fundamental interactions such as clicks to improve efficiency and minimize cascading \texttt{visit} command calls.

\heading{Interface specification:}
\texttt{visit} accepts an array of structured commands and executes them sequentially in a single call. Specifically, it supports four categories of commands:

\begin{myitemize}
  \item \textbf{Control access:} Navigate to the target control and perform a primitive interaction (i.e., click). 

  \textit{Target in the main tree:}
\begin{lstlisting}[style=tightjson]
{"id": "<target_id>"}
\end{lstlisting}

  \textit{Target in a shared subtree:}
\begin{lstlisting}[style=tightjson]
{"id": "<target_id>", 
"entry_ref_id": ["<ref_id>", "..."]}
\end{lstlisting}

  \item \textbf{Access-and-input text:} Access an \texttt{Edit} type control and input text:
\begin{lstlisting}[style=tightjson]
{"id": "<target_id>", "text": "<text>"}
\end{lstlisting}

  \item \textbf{Shortcut keys (auxiliary):} Issue a keyboard shortcut when a GUI click is insufficient (e.g., pressing ENTER to commit an edit):
\begin{lstlisting}[style=tightjson]
{"shortcut_key": "<key_combination>"}
\end{lstlisting}

  \item \textbf{FurtherQuery:} Request additional topology when the default core topology is insufficient. This command is exclusive and cannot be mixed with other commands in the same call. ``-1'' refers to fetching the entire forest.

\begin{lstlisting}[style=tightjson]
{"further_query": ["<node_id>", "..."]}
\end{lstlisting}
\end{myitemize}

\heading{Balancing efficiency and accuracy:}
\label{sec:visit_tradeoff}
We address the efficiency-accuracy tension in \texttt{visit} through three approaches.
First, \texttt{visit} bundles ``navigation'' and ``primitive interaction'' (e.g., click) into the \emph{atomic} \texttt{control-access} command and supports multiple commands per call (e.g., \texttt{visit([command1, command2])}),  enabling the LLM to execute several functions in one turn. Second, \texttt{visit} excludes complex interactions (i.e., actions supported by state declaration). When subsequent steps require composite interactions, \sys requires the LLM to stop emitting further visit commands. This forces the LLM to observe (e.g., verify the newly revealed dynamic controls after setting the scroll bar) before re-planning based on the actual UI state. \sys disallows mixing \texttt{visit} interface with other interfaces in the same turn to enforce this. An exception is \texttt{access-and-input-text}. We specifically include this command in the \texttt{visit} interface, which overlaps with state declaration, because the outcome of an ``edit'' action is limited and highly predictable, and therefore is  unnecessary to observe the execution result. This design improves efficiency by allowing the execution of \texttt{edit} commands with other \texttt{control-access} commands in a single turn, without a forced stop to observe. Third, \texttt{visit} supports shortcut commands for essential keyboard functions (e.g., pressing \texttt{ENTER} to commit text in an edit field). We instruct the LLM to use this judiciously.

\heading{Handling unstable UI interaction:}
We design several techniques to enhance the robustness of the executor.
First, we employ a fuzzy control matcher that combines control type, ancestor hierarchy, and name similarity when exact matching fails due to name variations or UIA's lack of guaranteed unique identifiers.
Second, we provide structured error feedback that explicitly describes control states and context to guide subsequent planning, e.g., when a target control is located but disabled.
Finally, we use a failure retry mechanism for GUI controls that may load slowly, retrying when deterministically expected controls are absent after interactions.
We exclude shortcut-key operations to prevent unintended side effects from repeated executions, e.g., pressing \texttt{Enter} twice.

\heading{Handling improper LLM instruction-following:}
Despite explicit instructions to output only target control identifiers, LLMs sometimes include navigational nodes. We address this by trusting the LLM's intended destination while ignoring its navigation process. The key insight is that functional nodes are topology leaves, while navigational nodes are non-leaves. We filter out non-leaf nodes, allowing the executor to retain only commands targeting functional nodes and handle navigation independently. This provides fault tolerance regardless of whether the LLM outputs navigational steps. Shortcut-key commands following filtered commands are also removed to maintain consistency.

\begin{table}[]
\caption{\textit{state} declaration and \textit{observation} declaration interfaces.
UIA defines 34 control patterns in total.
These interfaces are extensible. For example, \texttt{set\_texts} builds on \texttt{TextPattern}, \texttt{set\_toggle\_state} builds on \texttt{TogglePattern}, and \texttt{set\_expanded/set\_collapsed} builds on \texttt{ExpandCollapsePattern}.}
\label{tab:InteractOps}
\setlength{\tabcolsep}{3pt}        
\renewcommand{\arraystretch}{1.15}
\footnotesize                     
\begin{tabularx}{\columnwidth}{@{}l l >{\raggedright\arraybackslash}X@{}}
\toprule
\makecell[l]{Interface} &
\makecell[l]{Control Pattern} &
\makecell[l]{Description} \\
\midrule
\texttt{set\_scrollbar\_pos} & \texttt{Scroll} &
\makecell[l]{Set scrollbar position to \texttt{x}\%} \\
\texttt{select\_lines} & \texttt{Text} &
\makecell[l]{Select one (or contiguous) line(s)}\\
\texttt{select\_paragraphs} & \texttt{Text} &
\makecell[l]{Select one paragraph or a\\contiguous paragraph range} \\
\texttt{select\_controls} & \texttt{Select} &
\makecell[l]{Single or multi-select controls} \\
\texttt{get\_texts} & \texttt{Text\&Value} &
\makecell[l]{Retrieve a control's text} \\
\bottomrule
\end{tabularx}
\end{table}

\subsection{State and Observation Declaration: Interaction-related Interfaces}
\label{sec:InteractOps}

Triggering a control's functionality requires interacting with it. Beyond basic interactions (e.g., clicks), controls may require complex interactions involving coordinated keyboard and mouse actions (e.g., multi-select controls). Under the UIA framework, a control describes its available functionality through a finite set of control patterns (e.g., \texttt{TextPattern}, \texttt{ScrollPattern}, \texttt{SelectionPattern}). We leverage these control patterns and corresponding UIA interfaces to encapsulate control interactions, thereby providing a new layer of abstraction. In this way, we realize \textit{state} declaration and \textit{observation} declaration (see \autoref{sec:designoverview}).

\textit{state} and \textit{observation} declarations reduce dependency on
precise visual perception and simplify complex interaction.
For example, an Excel cell's full content may not be fully
visible and would require several clicks to reveal.
\sys ships with a set of pre-wrapped, high-value, and complex interactions, and can be customized to add additional UIA control patterns accordingly (see \autoref{tab:InteractOps}).

\heading{Supporting precise perception by default:}
When \sys interacts with applications, controls may expose dynamic data that cannot be modeled offline.
We adopt a ``passive + active'' design for \texttt{get\_texts()}. 
Specifically, before each LLM call, \texttt{get\_texts()} is invoked in passive mode on all \texttt{DataItem} controls, and a truncated, structured result is forwarded into the prompt. 
The passive mode reduces fine-grained visual parsing and saves round-trip, and items with empty values are coalesced for brevity. 
In cases where truncated results are insufficient, LLMs can call \texttt{get\_texts()} in active mode to retrieve the full content. 
The operation is implemented on UIA \texttt{TextPattern} and \texttt{ValuePattern} and generalizes to non-DataItem controls.

\heading{Separating control access and complex interactions:}
Control identifiers are used by interaction-related interfaces. 
However, to prevent mixing with \texttt{visit} interface within the same turn and preserve accuracy (see~\autoref{sec:visit_tradeoff}), the use of static IDs from the navigation topology is explicitly prohibited by interaction-related interfaces.
Instead, those interfaces can only operate on controls specified by their label from the current screen's accessibility tree.

\section{Implementation}
\sys comprises over 18K lines of Python code and leverages the pywinauto library~\cite{pywinauto} to exercise UIA~\footnote{\sys will be released at \url{https://github.com/dmi-interface/DMI}}.

\subsection{UNG Construction}
\label{sec:navmodel}
This section describes the methodology for constructing the UNG, which is formally defined in \autoref{sec:Path-Unambiguous}.

\heading{Control identifier synthesis:}
We require a control identifier to label each UI control as a node in the UNG.
Since UIA lacks guaranteed globally unique identifiers, we adopt an XPath-like control identifier:
\begin{center}
\texttt{primary\_id|\allowbreak control\_type|\allowbreak ancestor\_path}
\end{center}
\texttt{primary\_id} uses the UIA \texttt{automation\_id} (if empty, falling back to the control \texttt{name}, or \texttt{[Unnamed]}); \texttt{control\_type} specifies the UIA-defined type (e.g., \texttt{TabItem}); and \texttt{ancestor\_path} provides a slash-delimited sequence of UI tree ancestors.
We avoid index-based addressing since dynamic menus can shift indices unpredictably.

\heading{GUI ripping:}
The UNG models transition relationships between UI controls. The UNG is built via differential capture. Exploration proceeds with depth-first search (DFS). First, obtain the accessibility tree of the target application, from which a list of UI controls is captured. Then activate a candidate control (i.e., click) from the list and capture again. Newly revealed controls define navigation edges. New top-level or modal windows are detected via \texttt{process\_id} and window listeners. The exploration is fully automated.

\heading{Access blocklist:}
The DFS-based ripping requires the application to return to the prior state before exploring new branches. However, certain controls (i) trigger external transitions (e.g., an ``Account'' button opening a Web browser) or (ii) bring the application into states that cannot be exited using standard  commands like \texttt{Esc} or \texttt{Close}. While state consistency can be force-restored by restarting the application after every interaction \cite{guiripping}, this incurs a prohibitive time cost. To bypass this bottleneck, we adopt a \textbf{semi-automated} approach by implementing a manual blocklist for these controls. The configuration leverages prior application knowledge and can be verified by monitoring the modeling process. This represents a deliberate trade-off: necessary human intervention in exchange for a massive gain in exploration efficiency. Maintaining the blocklist constitutes most of the manual effort, but it can be automated with additional modeling time.

\heading{Context-aware exploration:} Some controls are implicitly state-dependent and are only visible under specific conditions (e.g., PowerPoint's ``Picture Format'' tab appears only when an image is selected). Accordingly, we implement a context manager. We manually instantiate representative objects (e.g., inserting an image or a text box into the slides) along with their associated context types (e.g., image, text box). The explorer traverses each context independently and merges the results into a unified topology. Although this manual setup represents a small fraction of the overall manual effort, automating it remains challenging. Omitting it would compromise the completeness of the UNG.

\heading{Root node initialization:}
A virtual root is introduced, and controls on the initial screen are attached as its children. If multiple \texttt{TabItem} controls exist and one is active by default, we associate otherwise unscoped controls on the initial screen with that active tab control to ensure they are indexable. The process is fully automated.

\subsection{Descriptions of Controls and Navigation}

\heading{Output schema:}
Each UI navigation tree and subtree is serialized as compact structured text:

\begin{center}
\texttt{name(type)(description)\_id[children]}
\end{center}
Parentheses mark optional fields; square brackets encode hierarchical nesting. The \texttt{name}, \texttt{type}, and \texttt{description} are drawn solely from UIA properties provided by the application.
The \texttt{id} is a unique, consecutive integer used for concise references (replacing verbose control IDs). The children are a comma-separated list of child control information.

\heading{Truncating descriptions:}
Control descriptions are selectively attached and truncated (by characters or tokens). The \texttt{full\_description} property is always included for control with key type (e.g., \texttt{Menu}, \texttt{TabItem}, \texttt{ComboBox}, \texttt{Group},  \texttt{Button}) when available. If multiple controls share a name and the group includes at least one key type, descriptions are applied to all. The \texttt{full\_description} property is preferred; otherwise, a location-based identifier is used (\autoref{sec:navmodel}). The non-leaf (navigational) nodes are fewer but pivotal; by default, full descriptions are included for these nodes when available.

\subsection{The \texttt{visit} Interface
}
\texttt{visit} receives a JSON array of commands from the LLM, and translates each into concrete GUI or keyboard actions.

\heading{Path resolution:}
The executor first discards commands targeting non-leaf (navigational) nodes and any following shortcut\_key commands, thereby retaining only the intended functional (leaf) targets. Each retained command is then resolved to a unique root-to-target navigation path.

\heading{Path navigation:}
The executor traverses this path from the current UI state. On each iteration, it fetches the topmost valid window and all descendant controls, and matches the parsed path from the end backward against the visible hierarchy. If no control from the remaining path exists in the current window, it closes that window. The ``closing'' follows a priority of OK > Close > Cancel, favoring the saving of modifications. Once a match is established, it proceeds forward along the path. Navigation robustness is enhanced by fuzzy matching and failure retries.

\heading{Control Interaction:}
After navigation, the executor performs the interaction based on the control access command, such as a click or a click followed by text input.

\subsection{The interaction-related Interfaces}

We adopt a conservative execution strategy for these interfaces. If any controls do not support the required pattern, the executor returns an error and does not partially execute. The executor returns a structured status (e.g., scrollbar positions).

\section{Evaluation}
\label{sec:evaluation}

Our key takeaways from the evaluation are:
\begin{myitemize}
  \item{Compared to the GUI-only baseline, \sys increases success rates by \textbf{1.67$\times$} and reduces steps by \textbf{43.5\%} over several models and tasks.}
  \item{\sys enables global planning and allows LLMs to complete user intent using a \textbf{single} call (in >61\% of cases).}
  \item{With \sys (GUI+\sys), most failures are \textbf{policy}-level (81\%, semantic planning); only 19\% are mechanism-level (navigation/interaction).}
\end{myitemize}

\subsection{Setup}

\heading{Case studies:}
We evaluate Microsoft Word, Excel, and PowerPoint as representative, feature-rich productivity applications. These applications collectively cover a range of scenarios, from text and tabular editing to graphics, and expose over 5,000 controls each.  Their complex UI features nested dialogs, child windows, and dynamic panes, with a navigation depth exceeding 10. This results in a complex navigation topology with cycles and numerous merge nodes. The suite broadly covers UIA control types and patterns, and mixes structured controls with substantial unstructured content (e.g., arbitrary document text, free-form shapes on slides).

\heading{Benchmark:}
We draw tasks from the widely used OSWorld benchmark \cite{osworld}; specifically, the OSWorld-W (Windows) portion comprising 27 single-app scenarios for PowerPoint, Excel, and Word. We evaluate on the full set of single-app scenarios. We exclude the multi-app subset because it exercises the operating system's controls. Although modeling the OS controls is feasible, it would have required a significant amount of time for modeling.

\heading{Baseline:}
We adopt Microsoft's Windows agent framework UFO-2 \cite{ufo2} as our baseline; it natively integrates UIA. For a fair comparison, we use the combination UFO2-base + action sequence (denoted UFO2-as). Here, UFO2-base is GUI-only. UFO-2 also offers Office-specific COM APIs, but those are not general and are therefore excluded. The action sequence mechanism reduces round trips by allowing the LLM to emit multiple actions in a single turn, provided all referenced controls are currently visible in the app UI. In the baseline (and all our settings), we register a  UIA event handler to trigger applications to expose full control trees (avoiding lazy loading artifacts). We also change control labeling: before calling the LLM, the baseline labels accessible-tree controls and passes the labels in the prompt; to distinguish these labels from our numeric IDs in the navigation topology, labels are alphabetic (e.g., ``A'', ``HF'').

\heading{Our approach:} On top of UFO2-as, we introduce our declarative \sys to evaluate its effectiveness. Prompts instruct the LLM to prefer \sys. When \sys cannot complete a step, the LLM may fall back to the baseline's imperative GUI primitives (e.g., click, drag\_on\_coordinates, keyboard\_input).

\heading{Methodology:}
Each task is capped at 30 steps to prevent excessive retries and is run three times, then averaged. Office experiments use Microsoft 365 builds. The LLMs are OpenAI GPT-5 and GPT-5-mini. The API variants are reasoning models with a configurable reasoning effort: minimal, low, medium (default), high. We compare GPT-5 at medium vs minimal to emulate ``reasoning'' vs ``non-reasoning'' modes. No fine-tuning is applied.

\subsection{Offline Phase: UI Navigation Modeling Cost}

\heading{Modeled graphs:}
We build the UNG with the prototype modeler and then transform it into a path-unambiguous forest. In the raw modeled graphs, Excel/PowerPoint/Word each exceeds 4K controls. To control context cost, we extract a core topology from the forest: Excel $\approx$ 2K, Word $\approx$ 1K, and PowerPoint $\approx$ 1K controls. Word and Excel take a forest; PowerPoint's core topology is a single tree. No nested references exist among shared subtrees.

\heading{Cost:}
Automated modeling per application takes < 3 hours. Some manual setup is required (see \autoref{sec:navmodel}). We assume the operator has reasonable app proficiency (e.g., can identify controls that jump to external applications or drastically reshape the UI, either a priori or by observing the modeling execution). Average manual effort is around 1.5 person-days per application. The resulting model is version-specific but reusable across machines for the same application build.

\begin{table}[t]
  \centering
  \caption{\small Results across interfaces and models.}
  \label{tab:evaltable}
  \resizebox{\columnwidth}{!}{%
  \begin{tabular}{l l l l c c c}
    \toprule
    Interface & Knowledge & Model & Reasoning & SR & Steps & Time(s) \\
    \midrule
    GUI-only & /           & GPT-5  & Medium & 44.4\% & 8.16 & 392 \\
    GUI-only & Nav.forest  & GPT-5  & Medium & 42.0\% & 8.41 & 353 \\
    \textbf{GUI+\sys}  & \textbf{Nav.forest}  & \textbf{GPT-5}  & \textbf{Medium} & \textbf{74.1\%} & \textbf{4.61} & \textbf{239} \\
    \specialrule{\lightrulewidth}{.25em}{.25em}
    \addlinespace
    \specialrule{\lightrulewidth}{.25em}{.25em}
    GUI-only & /           & GPT-5  & Minimal & 23.5\% & 8.42 & 251 \\
    \textbf{GUI+\sys}  & \textbf{Nav.forest}  & \textbf{GPT-5}  & \textbf{Minimal} & \textbf{40.7\%} & \textbf{5.52} & \textbf{140} \\
    \specialrule{\lightrulewidth}{.25em}{.25em}
    \addlinespace
    \specialrule{\lightrulewidth}{.25em}{.25em}
    GUI-only & /           & 5-mini & Medium & 17.3\% & 7.14 & 171 \\
    GUI-only & Nav.forest  & 5-mini & Medium & 23.5\% & 6.32 & \textbf{150} \\
    \textbf{GUI+\sys}  & \textbf{Nav.forest}  & \textbf{5-mini} & \textbf{Medium} & \textbf{43.2\%} & \textbf{4.43} & 167 \\
    \bottomrule
  \end{tabular}%
  }
\end{table}

\subsection{Online Phase: End-to-end Performance}
\label{sec:End-to-end Performance}
\begin{figure}[t]
  \centering
  \begin{subfigure}{0.99\linewidth}
    \centering
    \includegraphics[width=\linewidth,trim=0mm 2.5mm 4mm 6mm,clip]{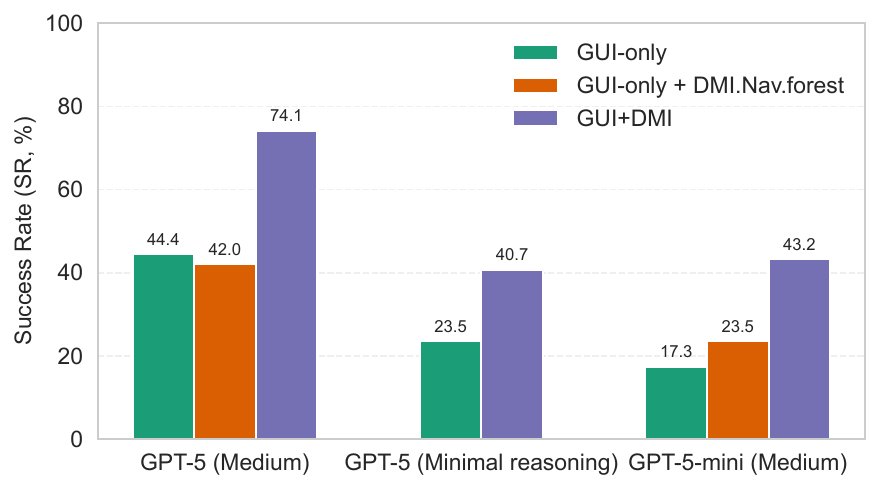}

    \caption{\small Success rate.}
    \label{fig:sub1}
  \end{subfigure}


  \begin{subfigure}{0.99\linewidth}
    \centering
    \includegraphics[width=\linewidth,trim=0mm 2.5mm 4mm 5mm,clip]{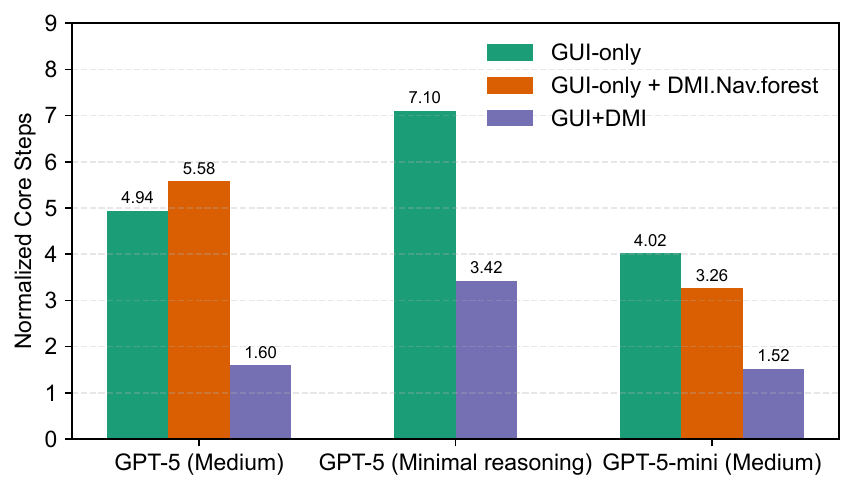}
    \caption{\small Normalized Core steps.}
    \label{fig:sub2}
  \end{subfigure}

  \caption{\small Visualization of performance evaluation. Core steps refer to the steps without the fixed 3-step agent framework overhead (see \autoref{sec:End-to-end Performance} One-shot task completion). Normalization refers to the intersection of tasks that all methods solve.}
  \label{fig:twopdfs}
\end{figure}

\heading{Terminology:}
Reasoning denotes the configured reasoning effect; SR is the average success rate. Step denotes the average number of LLM calls (i.e., round trips). Time is the average completion time. All metrics are calculated on successful cases only (\autoref{tab:evaltable}). We also report normalized steps (\autoref{fig:twopdfs}\subref{fig:sub2}), which are calculated based on the intersection of tasks successfully solved by all compared methods (GUI Baseline, GUI+\sys, and Ablation).

\heading{Settings:}
We compare three settings: (1) GPT-5, reasoning mode (core setting); (2) GPT-5, minimal reasoning; and (3) GPT-5-mini, reasoning mode.

\heading{Overall improvement:}
In the core setting, \sys yields substantial improvements over the baseline: raising success from 44.4\% to 74.1\% (1.67$\times$), cutting steps from 8.16 to 4.61 ($-$43.5\%), and reducing completion time by 39\%. Under minimal reasoning, success improves from 23.5\% to 40.7\%, steps drop from 8.42 to 5.52, and time declines from 251s to 140s. With GPT-5-mini, \sys yields an absolute +25.9\% success gain  (2.5$\times$ over GUI-only) and 38\% fewer steps; average time is comparable (171s baseline vs 167s \sys) because we report only successful runs, and GUI-only primarily succeeds on shorter/easier cases. \autoref{tab:evaltable} shows the overall results evaluated on the full set.

When normalized to the \textbf{intersection} of tasks solved by all methods (GUI-only, GUI+\sys, and Ablation), \sys demonstrates even greater efficiency gains. On this common subset, \sys reduces the normalized steps from 7.94 (GUI-only) and 8.58 (Ablation) to 4.60 (GUI+\sys) under the GPT-5 (medium reasoning) setting. This further confirms that \sys significantly reduces the interaction overhead (see \autoref{fig:twopdfs}\subref{fig:sub2}). It is also important to note that tasks solved without \sys (GUI-only) \textbf{remain solvable} with \sys (GUI+\sys) in our evaluation.

\heading{One-shot task completion:}
In the core setting, with \sys, over 61\% of successful trials are completed in 4 steps. The architecture of the UFO-2 framework features a multi-agent design where a HostAgent orchestrates the overall workflow while specialized AppAgents manage tasks for individual applications. UFO-2's workflow incurs a fixed 3-step overhead. These 4 steps correspond to:
(1) HostAgent decomposes the user task by application and opens/activates the target app.
(2) APPAgent executes the delegated subtask.
(3) APPAgent verifies the result and decides on handoff.
(4) HostAgent verifies overall task completion.
Thus, for single-app tasks, \sys enables the LLM to complete the core user intent in a single LLM call. 
Although the baseline supports action sequences, it is constrained to currently visible controls and cannot plan over controls that are not yet exposed. Declarative \sys permits global planning, which cuts the number of LLM calls substantially, as depicted in \autoref{fig:twopdfs}\subref{fig:sub2}.

\subsection{Overhead}

\heading{Token cost:}
Compared with the baseline, \sys introduces additional context tokens. Over 80\% of this overhead comes from the navigation forest; the remainder stems from the \sys usage prompt and the truncated \texttt{DataItem} payloads returned by \texttt{get\_texts()}. Empirically, each control contributes ~15 tokens on average (measured under the \texttt{o200k\_base} encoding). The core topologies add approximately 30K (Excel), 15K (Word), and 15K (PowerPoint) tokens. While this increases per-call context, modern LLMs provide ample windows (e.g., GPT-5 400K, o3 200K, Claude Sonnet 4 200K). More importantly, \sys reduces interaction rounds substantially, resulting in total token usage per task being lower than the baseline in the core scenario.

\heading{Per-step latency:}
With the small model, per-call latency is higher, but overall task completion time drops because \sys requires fewer steps. Per-call latency with \sys is similar to the baseline for large models in both reasoning and non-reasoning modes.

\subsection{Ablation Study}

To determine whether \sys's performance gains stem from its declarative interface or simply from providing a static knowledge base, we conducted an ablation study on both large and small models, as shown in \autoref{tab:evaltable} and \autoref{fig:twopdfs}.

For the baseline UFO2-as, we provided the \sys navigation forest in the prompt while disabling the declarative interface. This configuration yields no significant change in performance (SR 42\% vs. 44\% baseline; Steps 8.41 vs. 8.16, Normalized Steps 8.58 vs. 7.94), indicating that the declarative interface, not the static knowledge, is the primary source of \sys's effectiveness.
In contrast, the less capable GPT-5-mini model shows a modest improvement when given the forest alone (SR 23.5\% vs. 17.3\% baseline; Steps 6.32 vs. 7.14; Normalized Steps 6.26 vs. 7.02). This suggests that supplementary topology knowledge can be helpful for models with less general-purpose knowledge. Crucially, however, the full \sys setting produced much larger gains (SR 43.2\%; Steps 4.43; Normalized Steps 4.52), confirming that the interface design is the dominant performance driver. The average time is slightly lower as the baseline succeeds on short/easy cases.

\subsection{Failure Analysis}

\begin{figure}[t]

  \centering
  \begin{subfigure}{0.48\linewidth}
    \centering
    \includegraphics[
        width=\linewidth,
        trim=0mm 2.5mm 4mm 6mm,
        clip
    ]{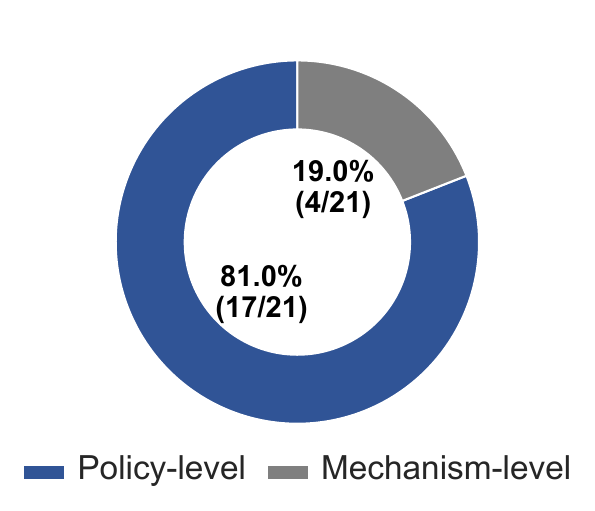}
    \caption{GUI\,+\,\sys}
    \label{fig:pie_gds}
  \end{subfigure}
  \hfill 
  \begin{subfigure}{0.48\linewidth}
    \centering
    \includegraphics[
        width=\linewidth,
        trim=0mm 2.5mm 4mm 6mm,
        clip
    ]{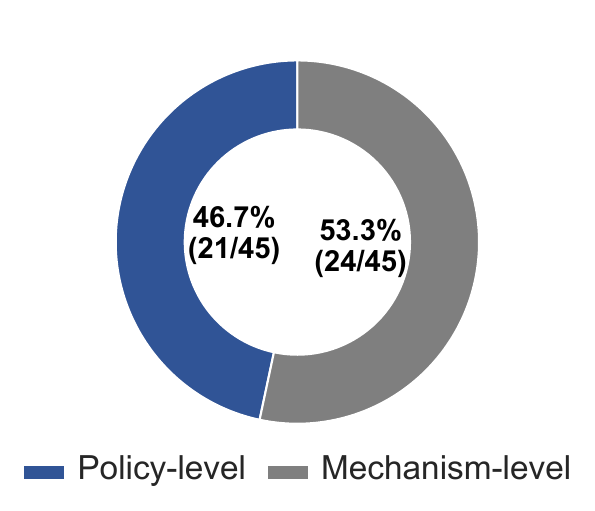}
    \caption{GUI-only Baseline}
    \label{fig:pie_gui}
  \end{subfigure}

  \caption{\small Failure-cause distribution (Policy/Mechanism) for \sys (GUI+\sys) and GUI-only baseline.}
  \label{fig:failure_pies}
\end{figure}

We analyze failures in the core setting and compare distributions between \sys and a GUI-only baseline (see \autoref{fig:failure_pies}). The analysis combines execution results with a chain-of-thought summary returned by the LLM. 

\heading{GUI+\sys (policy-centric failures):} The dominant causes are: ambiguous task descriptions (42.9\%); misinterpretation of control semantics (28.6\%; e.g., Word's ``subscript'' control in ``Find and Replace'' dialog applies to the whole Edit field rather than the selected range; Excel's rule of ``conditional formatting'' applies to all cells in the selected region, including blank cells); weak visual-semantic understanding (14.3\%); misunderstanding of subtle task semantics (9.5\%); and topology/modeling inaccuracies (4.8\%, see \autoref{sec:discussionandlimitations} (In)accurate navigation topology). 

Under the proposed method, errors concentrate at the policy level, specifically in task-semantic misunderstandings (52.4\%) and misperception of control functionality (28.6\%). This shift indicates that the declarative interface largely removes the mechanism-induced failures (navigation and interaction), leaving errors primarily in semantic understanding and planning, validating the intended decoupling. 

\heading{GUI-only baseline (mechanism-dominated failures):} The baseline shows many mechanism failures: control localization and navigation errors (14/45), and composite-interaction errors (7/45). It also shows additional task-semantic errors and misperception of control functionality (6/45). A further 18 failures overlap with the policy-centric categories above.
Because the LLM must split attention between policy and mechanism, more semantic mistakes appear. We observed similar errors in the ablation experiment, reinforcing that coupling policy with mechanism leads to avoidable mistakes.

\heading{Summary:}
The failure analysis pinpoints why \sys works: it cuts out mechanism-related errors (navigation and composite interaction) and re-centers policy, where LLMs are comparatively strong. Compared to GUI, \sys aligns better with LLM capabilities, substantially reducing reliance on precise vision and high-frequency interactions.

\subsection{Lessons Learned}

We discuss lessons learned when developing suitable declarative interfaces for LLMs. The following elements are necessary for \sys to achieve high success rates and efficiency.

\heading{Global unique identifier:}
A critical lesson is the need for applications to expose globally unique identifiers for controls. Although UIA recommends \texttt{AutomationId}, its global uniqueness is not enforced. As a result, current control identification methods like \texttt{XPath} are not reliable.

\heading{Rich control descriptions:}
Comprehensive and explicit descriptions in accessibility metadata are crucial for effective computer-use agents. For example, Excel's \texttt{Name Box} should explicitly state that pressing \texttt{Enter} is required to commit input. Future work can augment the textual navigation topology with descriptions synthesized from documentation or curated by LLMs.

\heading{Explicit navigation-node access:}
Our \texttt{visit} interface filters non-leaf (navigation) nodes, ensuring the interface takes over all navigation. If a task truly requires accessing a navigation node, the LLM can fall back to GUI operations. One alternative is to add an \texttt{enforced} parameter to control-access commands, allowing the caller to bypass filtering.

\section{Discussion and Limitations}
\label{sec:discussionandlimitations}

\heading{Completeness of \sys primitives:} The mechanism of GUI-use can be reduced to two parts: (1) making the target control visible on the screen, and (2) interacting with the control. \sys leverages the determinism of GUI navigation and abstracts control interactions into state representation to simplify complexity. It is important to note that \sys cannot simplify all complex interactions. For example, tasks like freely drawing a shape (e.g., canvas) or position-precise manipulation (e.g., figure adjustment) are inherently dependent on fine-grained, \textbf{non-standard} position-related operations. Such tasks are difficult to express in a stateful manner and cannot be easily simplified using a declarative approach.

Together, \sys and GUI form a fast-path/slow-path design. \sys is the fast path used by default; GUI is the slow-path fallback, invoked whenever (i) the needed interaction is outside \sys's coverage (e.g., freehand drawing) or (ii) execution hits exceptions/unexpected UI states (e.g., error pop-ups), ensuring functional completeness and recovery.

\heading{Generalization to other OSes:}
Our current implementation uses the Windows UI Automation (UIA) framework. We find that similar accessibility-based frameworks exist across OSes.
macOS provides \texttt{NSAccessibility} that supports traversing and invoking UI elements across applications. Its \texttt{role-subrole} hierarchies and associated attributes/actions map naturally to UIA control patterns.
Ubuntu exposes application widgets via \texttt{AT-SPI2}~\cite{Linux-AT-SPI2}. This mechanism also provides callable interfaces (e.g., Action, Value, Text, Selection) functionally equivalent to UIA control patterns.
Android \texttt{AccessibilityService} and \texttt{AccessibilityNodeInfo} provide comparable control pattern functionality.

\heading{Generalization to new applications:}
Extending \sys to new applications necessitates constructing an application-specific UNG to enable the access declaration based on navigation topology. As detailed in \autoref{sec:navmodel}, this involves manual configuration to facilitate the topology modeling.

\heading{Unsupported GUI applications:}
\sys is built atop Windows UI Automation (UIA), which also benefits many modern applications such as Google Chrome and Visual Studio Code. However, UIA does not cover all Windows applications.
Legacy applications from the Windows XP era often lack UIA support.
Real-time applications like video games and specialized graphics software often bypass standard UI widgets and may not expose accessible UIA trees, falling outside our methodology's scope.

\heading{(In)accurate navigation topology:}
Task success depends critically on accurate navigation topology modeling. Our prototype infers edges by comparing accessible trees before and after a click action, but cannot capture all dependencies.
For example, in Word's ``Find and Replace'' dialog, entering special text (e.g., +1, +2, +3) into the ``Rich Edit'' control dynamically renames the ``Next'' button to ``Go To''. Depth-first search exploration may miss such conditional changes, and without stable application identifiers, name changes break element matching. When topology is incomplete, LLMs can fall back to GUI-based imperative primitives, but may still fail due to poor visual localization of LLMs.

\heading{Handling dynamic controls:}
Applications might be driven by a dynamic UI that resists comprehensive pre-modeling. A canonical example is the Web browser. The main challenge of adapting \sys to Web browsers is the high volatility of Web controls—DOM structures and element attributes frequently change due to dynamic content rendering and continuous updates. This complicates navigation abstraction (access declaration) compared to relatively stable desktop applications. Handling dynamic controls is a challenging and promising direction for future work.
Nevertheless, \sys's state and observation declarations target UI controls interaction (e.g., set a scrollbar position, checkbox state, text in edit\_box), and thus can smoothly adapt to browsers.

\section{Related Work}

\heading{GUI-based large action models:}
UI-TARS \cite{uitars}, OS-ATLAS \cite{osatlas}, Aguvis \cite{aguvis}, and UGround \cite{uivground} train specialized large action models to navigate human-oriented GUI better. In contrast, \sys redesigns the interface to be LLM-friendly, eliminating the need for model training or fine-tuning.

\heading{Skill-based approaches:} 
AppAgentX \cite{appagentx} encapsulates UI sequences as skills. However, its reliance on visual element matching introduces a fundamental fragility. Similarly, AXIS \cite{apifirst} translates UI traces into application API calls, but its robustness is contingent upon the availability and stability of those APIs. In contrast, our accessibility-based interfaces offer a more robust approach. It operates without the need for app-specific APIs. Furthermore, our approach enables the composition of complex skills using declarative primitives, circumventing the need to rely on low-level UI sequences.

\heading{GUI ripping:}
Several studies in GUI testing model the application's structure to generate test cases. GUITAR \cite{guiripping} abstracts GUIs into event-flow graphs and synthesizes test cases from the model. Exploration strategies range from DFS to randomized search; platform-specific analogues include Android's monkey tool \cite{androidmonkey}. Broadly, some approaches \cite{testar, autodroidv1} model abstract state transitions, while others model reachability among GUI controls \cite{regui, autodroidv2}.  The key difference is how we treat path ambiguity at scale. Early Win32 GUITAR \cite{guiripping} expands multi-source graphs into multiple trees by cloning subgraphs to ensure unique paths; whereas modern, large UIs trigger exponential node growth. We instead introduce a cost-aware path disambiguation that yields a forest (namely, a main tree and shared subtrees) with controlled size.

\heading{Graph-based methods:}
AutoDroid-v1/v2 \cite{autodroidv1,autodroidv2} constructs UI/Element transition graphs and fine-tunes an on-device language model (LM) with sampled training data, which enables the LM to generate action sequences even when controls are not currently visible. However, this approach remains fundamentally imperative; the LM must plan both the high-level policy and the low-level mechanism, requiring it to enumerate fine-grained, step-by-step navigation sequences (e.g., \texttt{tap}(a) -> \texttt{tap}(b) -> \texttt{tap}(c)). When the LM's predicted next control is not found, the reachable path closest to that control is picked. However, this approach suffers from path ambiguity. For example, a single control labeled ``Color'' could correspond to different semantic functions (e.g., font color, outline color, or underline color). The greedy method does not reliably resolve such ambiguity. This poses a significant risk of incorrect functionality. In contrast, \sys provides unambiguous control access.

\heading{Novel system interfaces:} There is a long history of efforts to design new system interfaces.
Commutative interfaces \cite{commuter} enhance OS multi-core scalability.
SLO-aware interface \cite{MittOS} achieves millisecond tail tolerance.
\cite{LPN, PIX} proposes performance interfaces for systems.
Moreover, \cite{linux-api-study} examines modern Linux APIs' usage and compatibility, providing insights for new system interface design.

\heading{Other related work:}
SWE-agent \cite{SWE-agent} defines Agent-Computer Interface (ACI) to enhance software engineering. Agent S \cite{AgentS} proposes another ACI utilizing Accessibility APIs to label on-screen controls (similar to our baseline \cite{ufo2}). Agent S2 \cite {agents2} introduces the Mixture-of-Grounding that equips the LLM with specialized tools to enhance CUAs. SeeClick \cite{SeeClick} uses large vision-language models to locate controls.

\section{Conclusion}

OSes have long evolved interfaces to serve different users---from command-line interface (CLI) for experts to GUI for general users. LLM-powered agents are a fundamentally new class of OS users with distinct characteristics (large context-memory, reasoning) and constraints (weak visual capability, latency, token costs, imperfect instruction-following) that existing OS interfaces do not consider or address. As a result, LLMs underperform with GUI.

\sys addresses this gap by applying core systems principles (declarative over imperative, separating policy from mechanism, fast-path/slow-path) to build LLM-friendly OS interfaces.

\section*{Acknowledgments}

We sincerely thank the anonymous reviewers of EuroSys 2026 for their constructive comments, and our shepherd Jayson Boubin, for reviewing our camera-ready version.
We also thank Zhao Zhang and Zehao Song for proofreading drafts.
This work was supported in part by
the National Natural Science Foundation of China (No. 62502510).
Mingyu Li (\url{limingyu@ios.ac.cn}) and Haibo Chen (\url{haibochen@ios.ac.cn}) are the corresponding authors.

\balance
\bibliographystyle{ACM-Reference-Format}
\bibliography{reference}

\end{document}